\documentclass[prd,twocolumn,showpacs,groupedaddress,superscriptaddress,amsmath,amssymb]{revtex4-2} 
 
\usepackage[utf8]{inputenc}
\usepackage[english]{babel}	

\usepackage{stfloats}

\usepackage{cmap}
\usepackage{bbm}
\usepackage{fontenc}

\usepackage{enumerate}
\usepackage{multirow}
\usepackage{float}
\usepackage{comment}
\usepackage{lineno}

\usepackage{amsmath,amsfonts,amssymb,mathtools,yhmath}	
\usepackage{euscript} 
\usepackage{mathrsfs}
\usepackage{mathtext}
\DeclareUnicodeCharacter{2212}{-}

\usepackage{color}			
\usepackage{graphicx}
\usepackage[export]{adjustbox}
\DeclareGraphicsExtensions{.png,.jpg}
\graphicspath{}	
\usepackage[section]{placeins} 

\usepackage{physics}
\usepackage{braket}
\usepackage{xparse}	

\usepackage{slashed}
\usepackage{indentfirst} 
\usepackage{enumitem}

\usepackage[colorlinks=true, filecolor=blue, citecolor=blue,urlcolor=black]{hyperref}
\usepackage{orcidlink}
\usepackage{url}
\usepackage{cleveref}
\date{\today}

\begin{document}

\title{Probing Hidden Leptonic Scalar Portals using the NA64 Experiment at CERN}

\author{A.~Ponten\orcidlink{0009-0008-2463-2930}}
\affiliation{Uppsala University, Department of Physics and Astronomy, High Energy Physics, Ångströmlaboratoriet, Lägerhyddsvägen 1, 752 36 Uppsala, Sweden}
\author{H.~Sieber\orcidlink{0000-0003-1476-4258}}
\email[\textbf{e-mail}:]{henri.hugo.sieber@cern.ch}
\affiliation{ETH Z\"urich, Institute for Particle Physics and Astrophysics, CH-8093 Z\"urich, Switzerland}
\author{B.~Banto Oberhauser\orcidlink{0009-0006-4795-1008}}
\affiliation{ETH Z\"urich, Institute for Particle Physics and Astrophysics, CH-8093 Z\"urich, Switzerland}
\author{P.~Crivelli\orcidlink{0000-0001-5430-9394}}
\email[\textbf{e-mail}:]{paolo.crivelli@cern.ch}
\affiliation{ETH Z\"urich, Institute for Particle Physics and Astrophysics, CH-8093 Z\"urich, Switzerland}
\author{D. Kirpichnikov\orcidlink{0000-0002-7177-077X}}
\affiliation{Institute for Nuclear Research of the Russian Academy  of Sciences, 117312 Moscow, Russia}
\author{S.~N.~Gninenko\orcidlink{0000-0001-6495-7619}}
\affiliation{Institute for Nuclear Research of the Russian Academy of Sciences, 117312 Moscow, Russia}
\affiliation{Center for Theoretical and Experimental Particle Physics, Facultad de Ciencias Exactas, Universidad Andres Bello, Fernandez Concha 700, Santiago, Chile}
\author{M.~H\"osgen}
\affiliation{Universit\"at Bonn, Helmholtz-Institut f\"ur Strahlen-und Kernphysik, 53115 Bonn, Germany}
\author{L.~Molina Bueno\orcidlink{0000-0001-9720-9764}}
\email[\textbf{e-mail}:]{laura.molina.bueno@cern.ch}
\affiliation{Instituto de Fisica Corpuscular (CSIC/UV), Carrer del Catedratic Jose Beltran Martinez, 2, 46980 Paterna, Valencia, Spain}
\author{M.~Mongillo\orcidlink{0009-0000-7331-4076}}
\affiliation{ETH Z\"urich, Institute for Particle Physics and Astrophysics, CH-8093 Z\"urich, Switzerland}
\author{A.~Zhevlakov\orcidlink{0000-0002-7775-5917}}
\affiliation{Bogoliubov Laboratory of Theoretical Physics, JINR, 141980 Dubna, Russia}
\affiliation{Matrosov Institute for System Dynamics and  Control Theory SB RAS, \\  Lermontov str., 134, 664033, Irkutsk, Russia }

\begin{abstract}
In this study, we demonstrate the potential of the NA64 experiment at CERN SPS to search for New Physics processes involving $e\rightarrow\mu$ transitions after the collision of 100 GeV electrons with target nuclei. A new Dark Sector leptonic portal in which a scalar boson $\varphi$ could be produced in the lepton-flavor-changing bremsstrahlung-like reaction, $eN\rightarrow \mu N\varphi$, is used as benchmark process. In this work, we develop a realistic Monte Carlo simulation of the NA64 experimental setup implementing the differential and total production cross-section computed at exact tree-level and applying the Weisz\"{a}cker-Williams phase space approximation. Using this framework, we investigate the main background sources and calculate the expected sensitivity of the experiment. The results indicate that with minor setup optimization, NA64 can probe a large fraction of the available parameter space compatible with the muon $g-2$ anomaly and the Dark Matter relic predictions in the context of a new Dark Sector leptonic portal with $10^{11}$ EOT. This result paves the way to the exploration of lepton-flavour-changing transitions in NA64. 
\end{abstract}
\maketitle
\section{Introduction}
The known shreds of evidence for the existence of Dark Matter (DM) are based on its gravitational behaviour inferred from astronomical, astrophysical, and cosmological observations. Those notably include galaxy rotational curves \cite{Rubin:1970zza,Freeman:1970mx,1985ApJ...295..305V,Begeman:1991iy}, gravitational lensing \cite{Zwicky:1937zza,Tyson:1998vp,Clowe:2006eq}, Cosmic Microwave Background (CMB) anisotropies \cite{Planck:2013pxb,Planck:2018vyg}, or large-scale structure (LSS) formation \cite{Refregier:2003ct,2dFGRS:2005yhx,DES:2021wwk}, suggesting that the DM relic density abundance, $\Omega_\text{DM}\simeq0.27$, exceeds that of atoms by a factor $\Omega_\text{DM}/\Omega_\text{baryon}\simeq5$.\\ \indent
Nonetheless, fundamental questions remain open, such as the microscopic nature of DM or the origin of its observed relic density. Because of the absence of sizeable interactions with Standard Model (SM) particles, a popular framework, the so-called Dark Sector (DS), suggests that DM particles are neutral under SM forces but charged under a new force thus interacting with SM particles through mediators \cite{Kobzarev:1966qya,Blinnikov:1982eh,Foot:1991bp,Hodges:1993yb,Berezhiani:1995am}.\\ \indent 
Different types of particles can intertwin these interactions opening a broad parameter space and motivating new and complementary experimental avenues \cite{Jaeckel:2020dxj,Lanfranchi:2020crw}. Portal interactions involving lepton-flavor-changing (LFC) transitions mediated by a scalar particle could explain the origin of Dark Matter and at the same time accommodate the long-standing muon $g-2$ puzzle \cite{Gninenko:2022ttd}. In this case, a new scalar boson can act as a mediator between the two sectors carrying dark and leptonic quantum numbers. As a result, non-diagonal flavour interactions are allowed involving $e\leftrightarrow\mu$ transitions. The effective Lagrangian governing the interaction is
\begin{equation}
L_\text{int} = -h \bar{e}P_R\mu\varphi+h.c.,
\end{equation}
with $P_R = 1/2(1+\gamma_5)$, is the right-handed chiral projection operator, and $h$ is the coupling strength. Within this model, it is assumed that the new boson has a mass above the muon mass and would predominantly decay invisibly avoiding existing experimental constraints from muon decay experiments \cite{TWIST:2014ymv,Gninenko:2022ttd, Workman:2022ynf}. Given the thermally averaged annihilation cross-section, $\langle\sigma v\rangle\simeq\mathcal{O}(1\ \text{pb})$, the DM relic abundance associated with fermionic DS particles, $\psi_1$ and $\psi_2$, is given through \cite{Gninenko:2022ttd}
\begin{equation}
    \frac{h^2 g^2_{\varphi \psi_1 \psi_2} m_1^2}{4 \pi \left\{ m_\varphi^2 - \left( m_1 + m_2 \right)^2 \right\}^2} = \mathcal{O} \left(10 \, \text{pb} \right),
    \label{eq:DM-constraint}
\end{equation}
where $m_i$ is the mass of $\psi_i$ fermion and $g_{\phi \psi_1 \psi_2}$ is the coupling strength of the scalar boson $\varphi$ to the DS particles.\\ \indent
Besides the possibility of probing the particle nature of DM, such a model conveniently provides a solution to the muon anomalous magnetic moment $g-2$, which problem is associated to the discrepancy between the experimental measured value \cite{Muong-2:2023cdq} and the prediction from the Standard Model (SM) \cite{Davier:2017zfy,Keshavarzi:2018mgv,Colangelo:2018mtw,Hoferichter:2019mqg,Davier:2019can,Keshavarzi:2019abf,Kurz:2014wya,Melnikov:2003xd,Masjuan:2017tvw,Colangelo:2017fiz,Hoferichter:2018kwz,Gerardin:2019vio,Bijnens:2019ghy,Colangelo:2019uex,Colangelo:2014qya,Blum:2019ugy,Aoyama:2012wk,Aoyama:2019ryr,Gnendiger:2013pva,CMD-3:2023alj}. 
In this case, the contribution to the muon $g-2$ from a hypothetical scalar is:
\begin{equation}
    \Delta a_{\mu}^{\varphi} = \frac{h^2}{16 \pi^2} \frac{m_{\mu}^2}{m_{\varphi}^2} \int_0^1 dx \frac{x^2 \left( 1 - x \right)}{\left( 1 - x \right) \left( 1 - \frac{m_{\mu}^2}{m_{\varphi}^2} x \right) + \frac{m_{e}^2}{m_{\varphi}^2} x}.
    \label{eq:DLS_g2}
\end{equation}
where $m_e$ and $m_\mu$ are the masses of electron and muon respectively.\\
The potential of fixed target experiments to search for such a boson was described in \cite{Gninenko:2022ttd}. In this work, we focus on implementing such a model in a realistic \texttt{GEANT4}-based \cite{GEANT4:2002zbu,Allison:2016lfl} framework to derive the sensitivity of the NA64 experiment \cite{Gninenko:2013rka, NA64:2023wbi}. In addition, in this work,  the lepton flavour changing process $e(\mu)+N\rightarrow\mu(e)+N+\varphi$ is described through its differential and total production cross-sections. Those later are derived in section \ref{sec:production-cross-sections} both at exact tree-level (ETL) and in the Weisz\"{a}cker-Williams (WW) phase space approximation. The accuracies of the two approaches are compared through numerical studies in section \ref{sec:numerical-integration}. The underlying LFC physics is implemented in a fully realistic Monte Carlo (MC) simulation to study the potential reach of the NA64 experiment, as presented in section \ref{sec:search-at-na64}. We discuss in detail the signal yield estimate through the \texttt{GEANT4}-based \texttt{DMG4} \cite{Bondi:2021nfp, Oberhauser:2024ozf} package for the simulation of Dark Matter. Besides, we estimate the main background sources for a statistics of $10^{11}$ electrons on target (EOT). The final projected sensitivity for such a model is evaluated in section \ref{ref:sensitivity}. Finally, our results are summarised in section \ref{sec:conclusion}. 
\section{Cross-sections calculations\label{sec:production-cross-sections}}
In the following, the production cross-section is derived both at exact tree-level and using the Weisz\"{a}cker-Williams equivalent photon flux phase space approximation. We follow closely the notations of \cite{Liu:2016mqv,Liu:2017htz,Kirpichnikov:2021jev,Sieber:2023nkq}. For the sake of generality, the cross-sections are derived for generic incoming and outgoing leptons, denoted respectively by $l_i$ and $l_f$, with $l=e,\ \mu$.
\subsection{The exact tree-level}
The $\varphi$ boson production associated with the LFC bremsstrahlung-like reaction is defined by the kinematics
\begin{equation}
    l_i(p)+N(P_i)\rightarrow l_f(p')+N(P_f)+\varphi(k),
\end{equation}
where we defined $p=(E_i,\mathbf{p})$ and $p'=(E',\mathbf{p}')$ to be respectively the incoming and outgoing four-momentum of the initial and final state lepton, $k=(E_\varphi,\mathbf{k})$ is the $\varphi$ boson momentum, and $M$ the mass of the target (N). The nucleus has initial and final four-momenta $P_i=(M,\mathbf{0})$ and $P_f=(P_f^{0},\mathbf{P}_f)$, and is treated as a spineless boson to which the associated photon vertex rule reads
\begin{equation}
    \label{eq:nucleus-vertex}
    i e \mathcal{P}^{\mu} F(t) = i e ( P_{i}^{\mu} + P_{f}^{\mu} ) F(t),
\end{equation}
where $F^2(t) \equiv G_2^\text{el} (t)$ is the squared elastic form factor \cite{Liu:2017htz,Kirpichnikov:2021jev}
\begin{equation}
    F^2(t)= Z^2\left(\frac{a^2 t}{1 + a^2 t}\right)^2 \left( \frac{1}{1+t/d}\right)^2,
    \label{eq:form_factor_def}
\end{equation}
with $a= 111Z^{-1/3}/m_e$ and $d=0.164A^{-2/3}\ \text{GeV}^{-2}$ and $P_\mu = P_{i\mu }+ P_{f \mu}$. For the typical lead target properties of NA64 with $Z=82$ and $A=207$, the inelastic term contributing to Eq. \eqref{eq:form_factor_def} can be neglected \cite{Kirpichnikov:2021jev}. The relevant matrix element to the $2\rightarrow 3$ LFC process is then 
\begin{equation}    
    \label{eq:matrix-element-etl}
    \begin{split}
    i \mathcal{M}_{\varphi}^{2\rightarrow 3} &= i h e^2 \frac{F (q^2)}{q^2} \mathcal{P}_{\mu} \Bar{u}_{f}^{(s')} (p') \bigg\{P_{R} \frac{(\slashed{p'}+\slashed{k}) + m_{i}}{\tilde{s}} \gamma^{\mu}\\ 
    &+ \, \gamma^{\mu}  \frac{(\slashed{p}-\slashed{k}) + m_f}{\tilde{u}} P_{R}  \bigg\} u_{i}^{(s)} (p),
    \end{split}
\end{equation}
where $\bar{u}_f$ and $u_i$ are the spinors of the initial and final state leptons with spins $s,s'$, $P_R$ the right-handed projection operator and $q=P_i-P_f$ the momentum transfer to the nucleus. We also define the following Mandelstam variables
\begin{equation}
\begin{split}
\centering
    \label{eq:mandelstam_definition}
    \tilde{s} &= (p'+k)^2 - m_{i}^2, \\
    \tilde{u} &= (p - k)^2 - m_{f}^2, \\ 
    t_{2} &= (p'-p)^2  \\ 
    t &= -q^2, \\ 
    m_{\varphi}^2 &= \tilde{s} + t_{2} + \tilde{u} + t.
\end{split}
\end{equation}
where $m_i$ and $m_f$ are the leptons masses for either combinations of $(m_e,m_\mu)$. The average squared amplitude of the process is calculated using the \texttt{FeynCalc} package \cite{Mertig:1990an,Shtabovenko:2016sxi,Shtabovenko:2020gxv} from the Wolfram \texttt{Mathematica} program \cite{Mathematica}, such that from Eq. \eqref{eq:matrix-element-etl}
\begin{equation}
     \overline{|\mathcal{M}_{\varphi}^{2\rightarrow 3}|^{2}} = e^4 h^2 \frac{F^2(t)}{t^2} |\mathcal{A}_{\varphi}^{2\rightarrow 3}|^{2},
\end{equation}
where
\begin{equation}
    \begin{split}
    |\mathcal{A}_\varphi^{2 \rightarrow 3}|^{2} &= \frac{(\tilde{s}+\tilde{u})^2}{2\tilde{s}\tilde{u}}\mathcal{P}^2 - \frac{2t}{\tilde{s}\tilde{u}}(\mathcal{P}\cdot k)^2  \\ 
    + \frac{(\tilde{s}+\tilde{u})^2}{2\tilde{s}^2\tilde{u}^2}&\Delta{m^2}\bigg\{ \mathcal{P}^2t - 4\bigg(\frac{\tilde{u}(\mathcal{P}\cdot p) + \tilde{s}(\mathcal{P}\cdot p')
    }{\tilde{s}+\tilde{u}}\bigg)^2 \bigg\}.
    \label{eq:a2_ETL}
    \end{split}
\end{equation}
with $\Delta m^2 \equiv m_{\varphi}^2 - m_{i}^2 - m_{f}^2$ and 
\begin{equation}
\begin{split}
\centering
    \label{eq:mandelstam_definition_new}
    \mathcal{P}^2&=4M^2+t,\\
    \mathcal{P}\cdot p&=2ME_i-(\tilde{s}+t)/2,\\
    \mathcal{P}\cdot p'&=2M(E_i-E_\varphi)+(\tilde{u}+t)/2,\\
    \mathcal{P}\cdot k&=\mathcal{P}\cdot p-\mathcal{P}\cdot p'.
\end{split}
\end{equation}
The double-differential cross-section can then be expressed following \cite{Liu:2017htz,Davoudiasl:2021mjy}
\begin{equation}
    \label{eq:dsdxdcos_ETL}
    \begin{split}
     \left( \frac{d^2 \sigma_{2 \rightarrow 3}}{dx d \cos{\theta_{k}}} \right)_\text{ETL} = \frac{\alpha^2h^2 }{4 \pi} \frac{E_{i} |\textbf{k}|}{|\textbf{p}| V} &\int_{t_\text{min}}^{t_\text{max}} \, dt \, \frac{F^2 (t)}{t^2} \\ &\times \int_{0}^{2 \pi} \, \frac{d \phi_{q}}{2 \pi} \, \frac{ |A_\varphi^{2 \rightarrow 3}|^{2} }{ 8 M^2 }.       
    \end{split}
\end{equation}
with $\alpha$ the fine-structure constant, $t_\text{min}$ and $t_\text{max}$ the minimum and maximum momentum transfer respectively as in \cite{Liu:2017htz} and $x=E_\varphi/E_i$ the fractional energy transferred to $\phi$ dark scalar. The integration is performed in the reference frame where $\mathbf{V}=\mathbf{p}−\mathbf{k}$ is parallel to $\hat{z}$ axis and $\mathbf{k}$ is in the $\widehat{xz}$ plane. As such, $\phi_\mathbf{q}$ is the azimuthal angle of $\mathbf{q}$ defined in the polar coordinates system of the aforementioned reference frame (see \cite{Liu:2017htz} for more details). The $\theta_k$ is the radiation angle of the produced $\varphi$ scalar boson such that
\begin{equation}
    V = |\textbf{p} - \textbf{k}| = \sqrt{\textbf{p}^2+\textbf{k}^2 - 2 |\textbf{p}| |\textbf{k}| \cos{\theta_{k}}}.
\end{equation}
\subsection{The Weisz\"{a}cker-Williams approach\label{sec:ww-approximation}}
In this section, the production cross-sections are derived assuming that the incoming lepton energy, $E_i$, is much larger than the mass of its lepton, $m_i$, and of the scalar boson $m_\varphi$. As such we use the Weisz\"{a}cker-Williams approach and we closely follow the procedure of \cite{Kirpichnikov:2021jev,Sieber:2023nkq}. Therefore, we can define the equivalent photon flux from nuclear as 
\begin{equation}
    \chi^\text{WW} = \int_{t_\text{min}}^{t_\text{max}} \,dt\ \frac{t-t_\text{min}}{t^2}F(t)^2,
    \label{eq:chi_WW}
\end{equation}
where $t_{min}\simeq U^2 / 4 E_{i}^2 (1-x)^2$ and $t_{max}= m_{\varphi}^2 + m_{i}^2$. The definition of the function $U$ is given below in Eq. \eqref{eq:approx-mandelstam}. It is worth noting that Eq. \eqref{eq:chi_WW} can be expressed analytically as presented in \cite{Kirpichnikov:2021jev}
\begin{equation}
    \chi = Z^2 \left[ \tilde{\chi}(t_\text{max}) - \tilde{\chi}(t_\text{min}) \right].
\end{equation}
where we define $t_d = d$ and $t_a = 1/a^2$ and express the antiderivative as
\begin{equation}
\begin{split}
    &\tilde{\chi}(t) = \frac{t_d^2}{(t_a - t_d)^3} \bigg[ \frac{(t_a - t_d)(t_a + t_{min})}{t + t_a} \\ 
    &+ \frac{(t_a - t_d)(t_d + t_{min})}{t + t_d} + (t_a + t_d + 2t_{min}) \log{\frac{t + t_d}{t + t_a}} \bigg]. \\
    & \\
\end{split}
\end{equation}
Thus, the double differential cross-section in terms of $(x,\ \theta_k)$ can be expressed as \cite{Kirpichnikov:2021jev,Sieber:2023nkq}
\begin{equation}
    \label{eq:dsdxdcos_WW_raw}
    \left( \frac{d^2 \sigma_{2 \rightarrow 3}}{dx d \cos{\theta_{k}}} \right)_\text{WW} = \frac{\alpha \chi^\text{WW}}{\pi (1-x)} E_{i} |\textbf{k}|\frac{d \sigma_{2 \rightarrow 2}}{d(p \cdot k)} \biggr\rvert_{t=t_\text{min}},
\end{equation}
with it being evaluated at $t=t_\text{min}$. In this approximation, we can rewrite the Mandelstam variables in the form
\begin{equation}
    \label{eq:approx-mandelstam}
\begin{split}
    \centering
    \tilde{s} & \simeq -\frac{\tilde{u}}{(1-x)} \\
    t_{2} & \simeq \frac{x\tilde{u}}{(1-x)} + m_{\varphi}^2, \\
    U&=-\tilde{u},\\
    \tilde{u} & \simeq -x E_{i}^2 \theta_{k}^{2} - m_{\varphi}^2\frac{1-x}{x} + m_{i}^2 (1-x) - m_{f}^2,\\
    m_{\varphi}^2 & \simeq \tilde{s} + t_{2} + \tilde{u}, 
\end{split}
\end{equation}
where we used the definition of $\tilde{u}$ from Eq. \eqref{eq:mandelstam_definition} and only kept terms up to order $\mathcal{O}(\theta_{k}^2)$, $\mathcal{O}(m_{i}^2 / E^{\prime 2})$ and $\mathcal{O}(m_{i}^2 / x^2 E_{i}^2)$.\\ \indent
The differential cross-section on the right-hand side of Eq. \eqref{eq:dsdxdcos_WW_raw} for the $2\rightarrow2$ process can be expressed as \cite{Bjorken:2009mm}
\begin{equation}
\label{eq:2-2_cross_section}
    \frac{d \sigma_{2 \rightarrow 2}}{d(p \cdot k)} = 2 \frac{d \sigma_{2 \rightarrow 2}}{d t_2}
    = \frac{\alpha h^2}{2 \tilde{s}^2} |\mathcal{A}_\varphi^{2 \rightarrow 2}|^{2}.
\end{equation}
In particular, the spin-averaged amplitude squared, $|\mathcal{A}_\varphi^{2 \rightarrow 2}|^{2}$, for $l_i +\gamma \rightarrow l_f +a$ of the process reads
\begin{equation}
    \begin{split}
    \label{eq:a2_WW}
    |\mathcal{A}_\varphi^{2 \rightarrow 2}|^{2} &= -\frac{\left(\tilde{s}+\tilde{u}\right)^2}{2\tilde{s}\tilde{u}} \\
    &+ \Delta m^{2} \left\{ \frac{(\tilde{s} + \tilde{u})(\tilde{s} m_{f}^2 + \tilde{u}m_{i}^2)}{\tilde{s}^2\tilde{u}^2} - \frac{t_{2}}{\tilde{s}\tilde{u}} \right\}.
    \end{split}
\end{equation}
Evaluated at $t=t_\text{min}$, the above amplitude can be written as 
\begin{equation}
    \begin{split}
    &|\mathcal{A}_\varphi^{2 \rightarrow 2}|^{2}_{t_\text{min}} = \frac{x^2}{2(1-x)} \\
    &+ \Delta m^2 \cdot \frac{ \tilde{u}x + m_{\varphi}^2 (1-x) + x \left [ m_{f}^2+m_{i}^2(x-1) \right ] }{\tilde{u}^2}.\\
    \end{split}
    \label{eq:a2tmin}
\end{equation}
Plugging Eqs. \eqref{eq:a2tmin} and \eqref{eq:2-2_cross_section} into Eq. \eqref{eq:dsdxdcos_WW_raw} gives us the following result
\begin{equation}
    \begin{split}
    \left( \frac{d^2 \sigma_{2 \rightarrow 3}}{dx d \cos{\theta_{k}}} \right)_\text{WW}
    &= \frac{\alpha^2 h^2 }{2 \pi}\chi^\text{WW} E_{i}^2 x \beta_{k} \frac{(1-x)}{\tilde{u}^2} |\mathcal{A}_\varphi^{2 \rightarrow 2}|^{2}_{t_\text{min}},
    \end{split}
\end{equation}
where $\beta_k=\sqrt{1-(m_{\phi}^2/(x^2E_\varphi})$. Purely in terms of the fractional energy $x$, one writes
\begin{equation}
    \begin{split}
    \left( \frac{d \sigma}{dx}\right)_\text{WW} &= \frac{h^2 \alpha^2}{2 \pi} E_{i}^2 x \beta_{k} (1-x)\\
    &\times\int_{0}^{\theta_\text{max}} \,d \cos\theta_k\ \frac{|\mathcal{A}_\varphi^{2 \rightarrow 2}|^{2}_{t_\text{min}}}{\tilde{u}^2}\chi^\text{WW}.\\
    \end{split}
    \label{eq:dsdx_WW}
\end{equation}
The value maximum angle $\theta_\text{max}$ is estimated from numerical studies and is found to be $\theta_\text{max}\simeq0.1$, given that larger bound values do not contribute much to the integral. This result is in good agreement with previous works such as \cite{Kirpichnikov:2021jev,Sieber:2023nkq}.
\subsubsection*{Final state lepton kinematics}
In the case where the kinematics of the final state lepton are of interest, it is instructive to perform similar computations based on its fractional energy and emission angle, $(y,\ \psi)$, with $y=E_f/E_i$. The corresponding double-differential cross-section then reads \cite{Kirpichnikov:2021jev}
\begin{equation}
    \left( \frac{d^2 \sigma}{d \cos{\psi} d y} \right)_\text{WW} = \frac{\alpha \chi}{\pi} \frac{E_{i} |\textbf{p}'| }{1-y} \cdot\frac{d \sigma_{2 \rightarrow 2}}{d (p \cdot p')}\biggr\rvert_{t=t_\text{min}},
    \label{eq:dsdydpsi_WW_raw}
\end{equation}
for which we can express the Mandelstam variables in a similar fashion as in Eqs. \eqref{eq:approx-mandelstam}
\begin{equation}
    \label{eq:dsdy_mandelstam}
    \begin{split}
    t_{2} & \approx - y E_{i}^2 \psi^2 - m_{f}^2 \left( \frac{1-y}{y} \right) + m_{i}^2 (1 - y), \\
    \tilde{t} & \approx m_{\varphi}^2 - t_{2}, \\
    \tilde{s} & \approx \frac{\tilde{t}}{1-y},\\
    \tilde{u} & \approx - \frac{y\tilde{t}}{1-y}.
    \end{split}
\end{equation}
Plugging Eq. \eqref{eq:dsdy_mandelstam} into Eq. \eqref{eq:a2_WW} gives
\begin{equation}
\label{eq:amplitude-squared-lepton}
\begin{split}
    |\mathcal{A}_\varphi^{2 \rightarrow 2}|^{2}_{t_\text{min}} &= (1-y)^2\\
    \times\bigg\{ \frac{1}{2 y} &- \frac{\Delta m^2}{y t} + \frac{\Delta m^2 [ m_{f}^2 + y ( m_{i}^2 y + \Delta m^2) ] }{y^2 t^2} \bigg\}.
\end{split}
\end{equation}
Combining Eqs. \eqref{eq:2-2_cross_section} and \eqref{eq:amplitude-squared-lepton}, the double-differential cross-section of Eq. \eqref{eq:dsdydpsi_WW_raw} reads
\begin{equation}
    \label{eq:dsdydpsi_WW}
    \begin{split}
    \left( \frac{d^2 \sigma}{d \cos{\psi} d y} \right)_\text{WW} &= \frac{\alpha^2 h^2}{2\pi} E_{i}^2 \beta_{f} \chi^\text{WW} (1-y)^3 \\ \times \bigg( \frac{1}{2 \tilde{t}^2} - \frac{\Delta m^2}{\tilde{t}^3} &+ \frac{\Delta m^2 [ m_{f}^2 + y ( m_{i}^2 y + \Delta m^2) ] }{y \tilde{t}^4}  \bigg),
    \end{split}
\end{equation}
where $\beta_{f} = (1 - m_{f}^2/E_f^2)^{1/2} $.
\subsection{Numerical integration\label{sec:numerical-integration}}
\begin{figure}[H]
    \centering
    \includegraphics[width=0.4\textwidth]{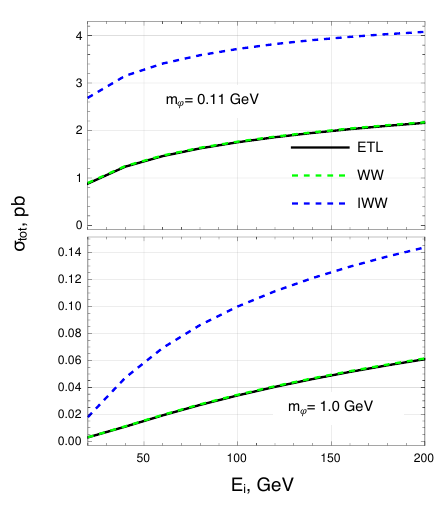}
    \caption{Total cross-sections evaluated at ETL (black line) and in both the WW and IWW approaches (respectively green and blue dash lines) for different initial electron energy. Top: $m_\varphi=110$ MeV. Bottom: $m_\varphi=1$ GeV.}
    \label{fig:total-etl-ww-iww-electron}
\end{figure}

The accuracy of the phase space approximation in the WW approach is estimated through comparison with the ETL computations and shown in Fig. \ref{fig:dsdx-dsdpsi-electron}. 
\begin{widetext}
\par\smallskip\noindent
\centerline{\begin{minipage}{\linewidth}
\begin{figure}[H]
    \centering
    \includegraphics[width=1.0\textwidth]{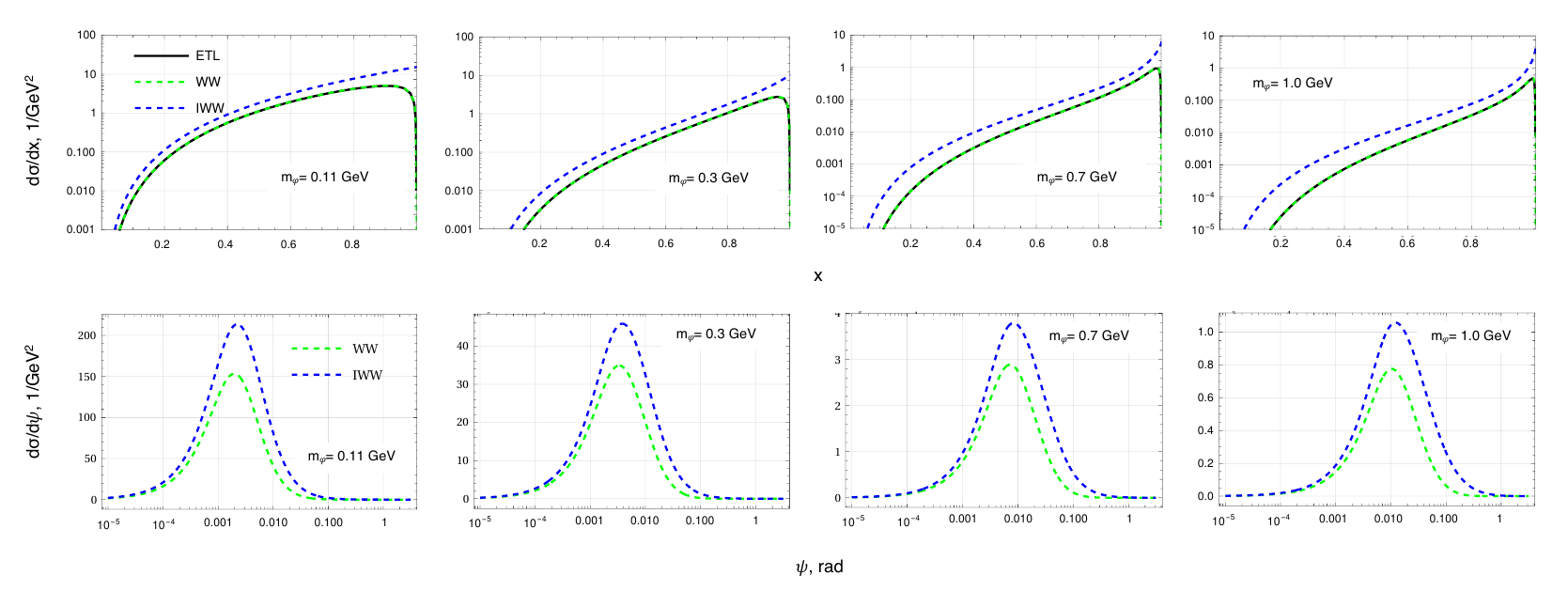}
    \caption{Top: Single-differential cross-sections as a function of the fractional scalar boson $\varphi$ energy, evaluated at ETL (black line) and in both the WW and IWW approaches (respectively green and blue dash lines). Eq. \eqref{eq:dsdxdcos_ETL} is evaluated with the \texttt{Mathematica} program \cite{Mathematica}. Bottom: Single-differential cross-section as a function of the outgoing lepton emission angle, $\psi$, in both the WW and IWW approaches. The mass range of the scalar boson spans from the $m_\varphi\geq m_\mu\simeq110$ MeV up to 1 GeV, with $h=10^{-4}$.}
    \label{fig:dsdx-dsdpsi-electron}
\end{figure}
\end{minipage}}
\end{widetext}
The relevant expressions, Eqs. \eqref{eq:dsdxdcos_ETL} and \eqref{eq:dsdx_WW}, are integrated numerically using methods from \texttt{Mathematica}. The primary lepton is chosen to be an electron with energy $E_i=100$ GeV. The mass range of the $\varphi$ boson spans from a few hundred MeV up to 1 GeV, the lower bound being justified by the kinematics requiring $m_\varphi>m_\mu$. For completeness, the improved Weisz\"{a}cker-Williams (IWW) approach is also shown, as it provides faster run-time performance to the cost of loss of accuracy to the ETL \cite{Liu:2017htz,Kirpichnikov:2021jev} (see Appendix \ref{app:iww-approach}). The relative error, expressed as $(\mathcal{O}_\text{exact}-\mathcal{O}_\text{approx.})/\mathcal{O}_\text{exact}$, is computed for both the IWW and WW approximations. In the WW approach, for the whole $m_\varphi$ range, an error of $\lesssim2\%$ is found with respect to the ETL. In the IWW case, the relative error is significant, in particular in the boundary regions, $x\rightarrow0$ and $x\rightarrow1$, due to the flux integral simplifications. \\ \indent
Because of the importance of the final-state lepton kinematics in the LFC process $e\rightarrow\mu$, the scattered angle of the muon is also shown in Fig. \ref{fig:dsdx-dsdpsi-electron} as a function of the $m_\varphi$ mass. It is found that in both approximations the differential cross-sections peak at $\psi\sim m_\varphi/E_i$. Because the ETL computations were not performed for the $(y,\ \psi)$ kinematical variables, the IWW and WW approaches are not compared to them.\\ \indent
For completeness, the total cross-section as a function of the incoming electron energy is shown in Fig. \ref{fig:total-etl-ww-iww-electron} for two selected scalar boson masses, $m_\varphi=110$ MeV and $m_\varphi=1$ GeV. While the WW reproduces accurately the ETL with a precision of $\mathcal{O}(\leq2\%)$, the IWW approximation overestimates at typical NA64 electron beam energy ($\sim100$ GeV) the tree-level computations by a factor $\sim2$ and $\sim3$ respectively for the lower and higher mass choices.\\ 
\section{Search strategy at NA64\label{sec:search-at-na64}}
\subsection{Experimental setup}
NA64 is a fixed target experiment at the CERN Super Proton Synchrotron (SPS) accelerator searching for Dark Sectors in the collisions of high energy lepton beams with an active target \cite{Gninenko:2013rka}. The experiment runs in two modes, NA64e \cite{NA64:2016oww,NA64:2023wbi,NA64:2023ehh} and NA64$\mu$ \cite{Gninenko:2014pea,Sieber:2021fue}, with different detector setups, using the 100 GeV H4 beamline with electrons/positrons and the 160 GeV M2 muon beam respectively. Since $\varphi$ mediates a non-diagonal interaction between a muon and an electron, it could be produced in both NA64 configurations via the bremsstrahlung-like process outlined in section \ref{sec:production-cross-sections}. This paper focuses only on the sensitivity to $\varphi$ production using the H4 beamline.\\ \indent
The experimental technique for the search of $\varphi$ in the collision $e^-N\rightarrow \mu N\varphi$ relies on the missing energy technique developed by NA64 since $\varphi$ decays invisibly. Single $e^-$ originating from the slow extracted beam are tagged through a set of scintillator and veto counters. A magnet spectrometer consisting of two dipole magnets and tracking detectors (Micromegas chambers (MMs), straw tubes (ST) and gas electron multiplier (GEM)), measures the incoming momentum with a precision of $\delta p/p\simeq1\%$ \cite{Banerjee:2015eno}. The hadron contamination at the H4 beamline with 100 GeV electrons is at the level of $\pi/e^-< 10^{-2}$, with a $\sim3\%$ kaon contamination of the pion rate \cite{Andreev:2023xmj}. The synchrotron radiation detector (SRD) identifies the electrons over the remaining hadrons. The SRD consists of an array of a lead-scintillator sandwich calorimeter of a fine segmentation placed after the bending magnet. It measures the synchrotron radiation of the incoming high energy electrons through the magnetic field \cite{Depero:2017mrr}. A 95$\%$ electron identification is achieved, thus suppressing the remaining hadron contamination in the H4 beam down to a level of $\sim 2\times 10^{-5}$ \cite{NA64:2023wbi}. The electrons collide with the active target, a 40 radiation length ($X_0$) high-efficiency shashlik electromagnetic calorimeter (ECAL) made by sandwiched lead-scintillator plates where $\varphi$ dark scalar would be produced. A large high-efficiency VETO counter and three 7.5 nuclear interaction length ($\lambda_\text{int}$) iron hadronic calorimeters (HCALs) complete the setup hermeticity, detecting charged and neutral secondaries produced from the electron interaction in the target and  measuring any energy leakage.\\ \indent
A key ingredient of this search is the addition of a second spectrometer consisting of an MBPL dipole magnet together with four MMs trackers placed after the last HCAL of the NA64 setup. A signal-like event will be missing energy carried away by the $\varphi$ produced in the target (see Fig. \ref{fig:dsdx-dsdpsi-electron}) and a muon in the final state. Such muon would need to traverse the three HCALs, leaving $\sim2.5$ GeV per module corresponding to the energy of a minimum ionizing particle (MIP). The magnet spectrometer will allow us to identify it and reconstruct its momenta. The missing energy is measured as the difference between the incoming beam electron's energy and the total energy carried away by the final muon: 
\begin{equation}
    E_\text{miss} = E_e^i - E_\text{ECAL}^f - E_\text{HCAL}^f - E_{\mu}^f.
    \label{eq:missing-energy}
\end{equation}
$E_e^i$ refers to the initial electron energy, $E_\text{HCAL}^f$ and $E_\text{ECAL}^f$ are the energy in the HCAL and ECAL calorimeters respectively, and $E_{\mu}^f$ is the energy of the final state muon reconstructed in the magnet spectrometer.\\ \indent
A schematic representation of the NA64 setup focusing on the part after the target is shown in Fig. \ref{fig:detector-setup}.\\ \indent
\begin{widetext}
\par\smallskip\noindent
\centerline{\begin{minipage}{\linewidth}
\begin{figure}[H]
    \centering
   \includegraphics[width=0.85\textwidth]{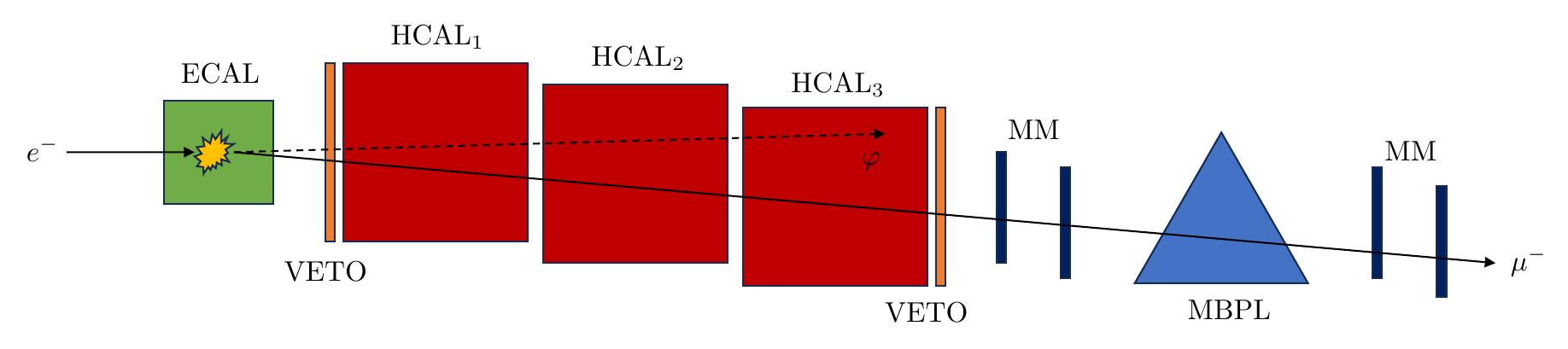}
    \caption{Schematic view of the setup at NA64e optimized for $ e \rightarrow \mu + \varphi $ searches. The MBPL magnet combined with four MM trackers constitute the magnet spectrometer. The whole set-up is not shown in the diagram focusing on the relevant part for the $e$ to $\mu$ process. A detailed description of the full setup can be found in \cite{NA64:2023wbi}.}
    \label{fig:detector-setup}
\end{figure}
\end{minipage}}
\end{widetext}
\subsection{Signal yield estimates\label{sec:signal-yield}}
\begin{figure}[h]
    \centering
    \includegraphics[width=0.38\textwidth]{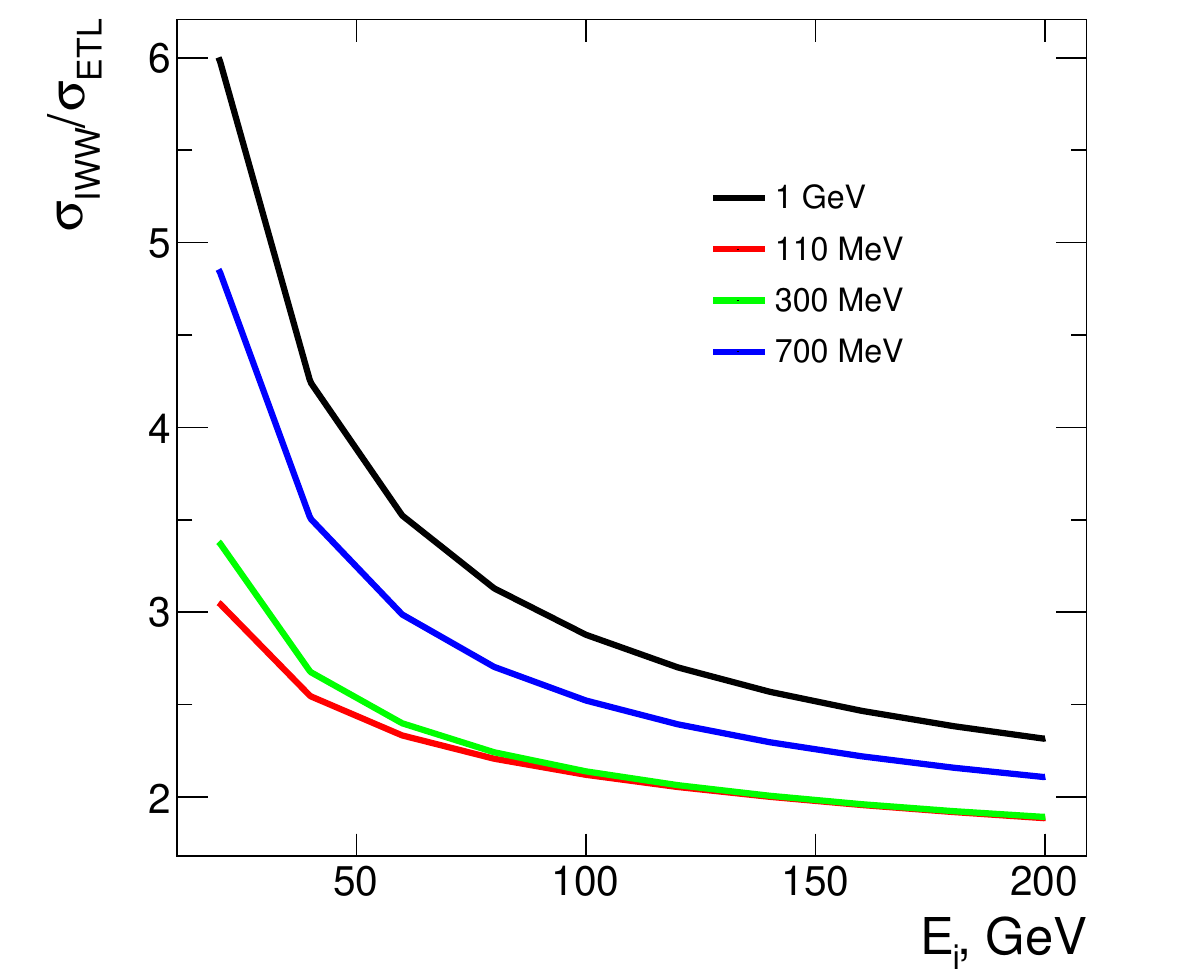}
    \caption{Ratios of the IWW approximated total cross-section to the ETL value for different initial electron energies and mass values.}
    \label{fig:k-factors}
\end{figure}

\begin{figure}[H]
    \centering
    \includegraphics[width = 0.38\textwidth]{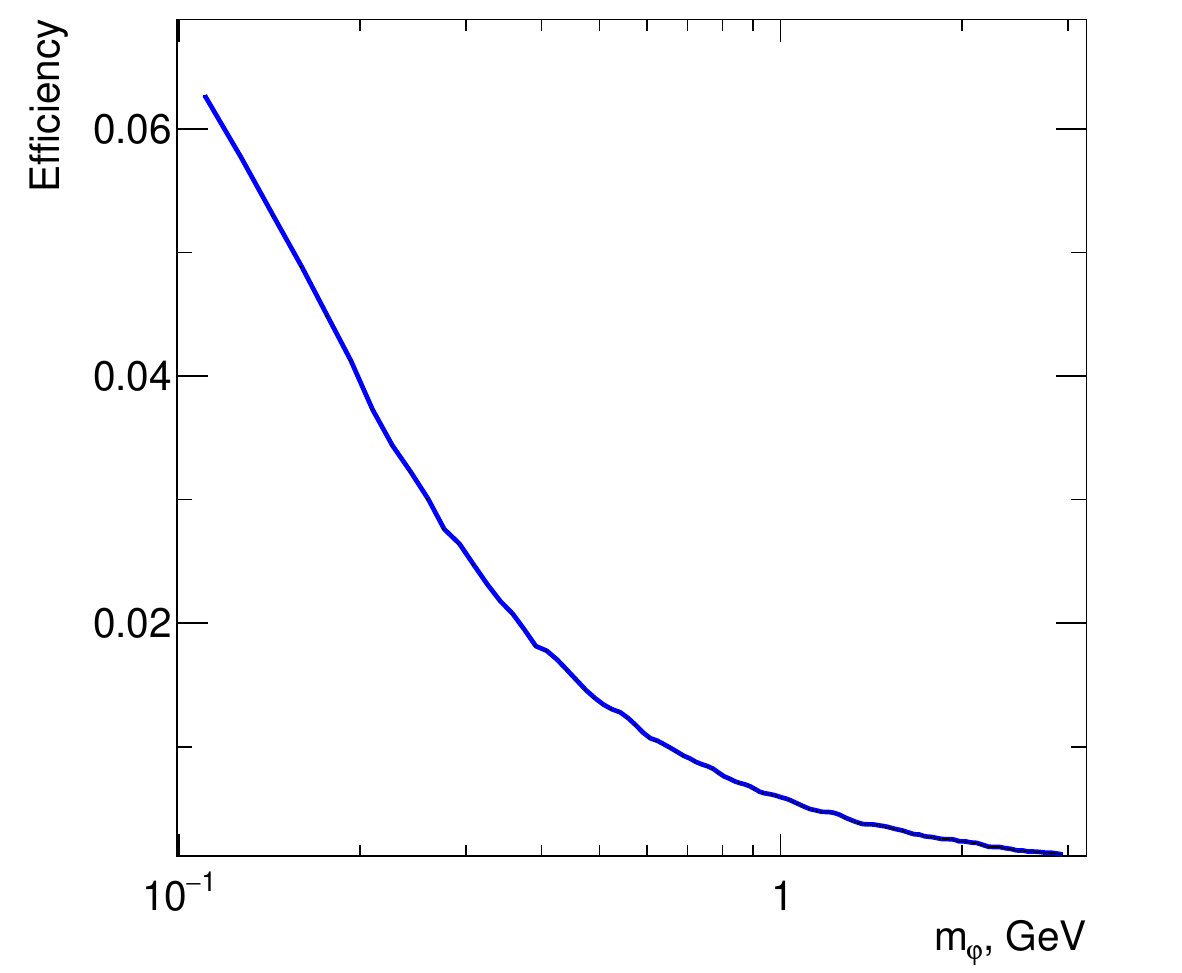}
    \caption{Signal selection efficiency with respect to the mass of $\varphi$. As the mass increases, the final state muon is likely to carry less of the incoming electron energy (see Fig. \ref{fig:dsdx-dsdpsi-electron}), increasing the likelihood of the muon to be absorbed in the ECAL or HCALs and not be reconstructed in the magnet spectrometer.}
    \label{fig:efficiency}
\end{figure}

For a realistic estimate of the NA64 experiment sensitivity to the scalar boson from the $e\rightarrow\mu$ LFC bremsstrahlung-like process, the full physics of the underlying model is implemented in the MC using the setup illustrated in Fig. \ref{fig:detector-setup}. In particular, for the signal simulation, we use the \texttt{DMG4} package \cite{Bondi:2021nfp, Oberhauser:2024ozf} as it provides an application programming interface to \texttt{GEANT4} \cite{GEANT4:2002zbu,Allison:2016lfl}. The double- and single-differential cross-sections derived in section \ref{sec:production-cross-sections} are implemented in the MC simulation framework. The related distributions are used to perform the sampling of the final states' kinematics using a Von Neumann-based accept-reject method \cite{UBHD66609337}. The mean free path of the interaction is estimated at run-time for each particle step within the target based on pre-computed tabulated ratios of the IWW approximation to the ETL value of the total cross-section (see Fig. \ref{fig:k-factors}). This is done for a wide range of $m_\varphi$ masses and initial electron energies.\\ \indent
Having implemented the underlying production mechanism in the simulations, the total number of expected signal events, $\mathcal{N}_\varphi$, is given by
\begin{equation}
    \mathcal{N}_\varphi=N_\text{EOT}\times\frac{\rho\mathcal{N}_A}{A}\sum_{k}\sigma_{e\rightarrow \varphi\mu}(E_k)\Delta L_k,
\end{equation}
where $N_\text{EOT}$ is the number of electrons on target (EOT), $\rho$ and $A$ the target density and atomic weight, $\mathcal{N}_A$ the Avogadro number, $\sigma_{e\rightarrow\varphi\mu}$ the production cross-section at the $k-$step within the target evaluated at an electron energy $E_{k}$ and $\Delta L_k$ the step length. Note that the total cross-section depends also on the values of the mass of the $\varphi$ boson and coupling $h$.\\ \indent
Based on the simulation results, the signature of $\varphi$ in the experimental setup was characterized. Signal-like events were identified according to the following criteria: (i) One and only one particle compatible with the trajectory expected for a $\mu^-$ reconstructed in the magnet spectrometer. (ii) The energy deposit in each HCAL module and in the VETO is consistent with that of a single minimum ionizing particle. The minimum required energy for a muon to pass all three HCALs at NA64 is approximately 7.5 GeV. (iii) A missing energy larger than 20 GeV (see Eq. \eqref{eq:missing-energy}). This value was chosen to optimize the ratio of signal over background, as it will be explained in more detail in the next sections. \\ \indent
 Since $\varphi$ is likely to carry most of the initial electron's energy (see Fig. \ref{fig:dsdx-dsdpsi-electron}), many final state muons have too little energy to overcome the calorimeters, and are either fully absorbed in the ECAL or stopped by the HCAL modules. As the mass $m_\varphi$ increases, the differential cross section is increasingly peaked towards $x = 1$ (Fig. \ref{fig:dsdx-dsdpsi-electron}), meaning even fewer signal muons are energetic enough to penetrate all HCALs. To be selected as candidate signal events, the muons produced in the interaction will exit the last HCAL, and will pass through the magnet gap hitting all MM trackers in the magnet spectrometer. The signal efficiency as a function of the mass ($m_\varphi$), is shown in Fig. \ref{fig:efficiency} after applying the selection cuts (i-iii) described previously. 
\subsection{Potential background sources\label{sec:background-sources}}
\begin{figure}[H]
    \centering
    \includegraphics[width = 0.5\textwidth]{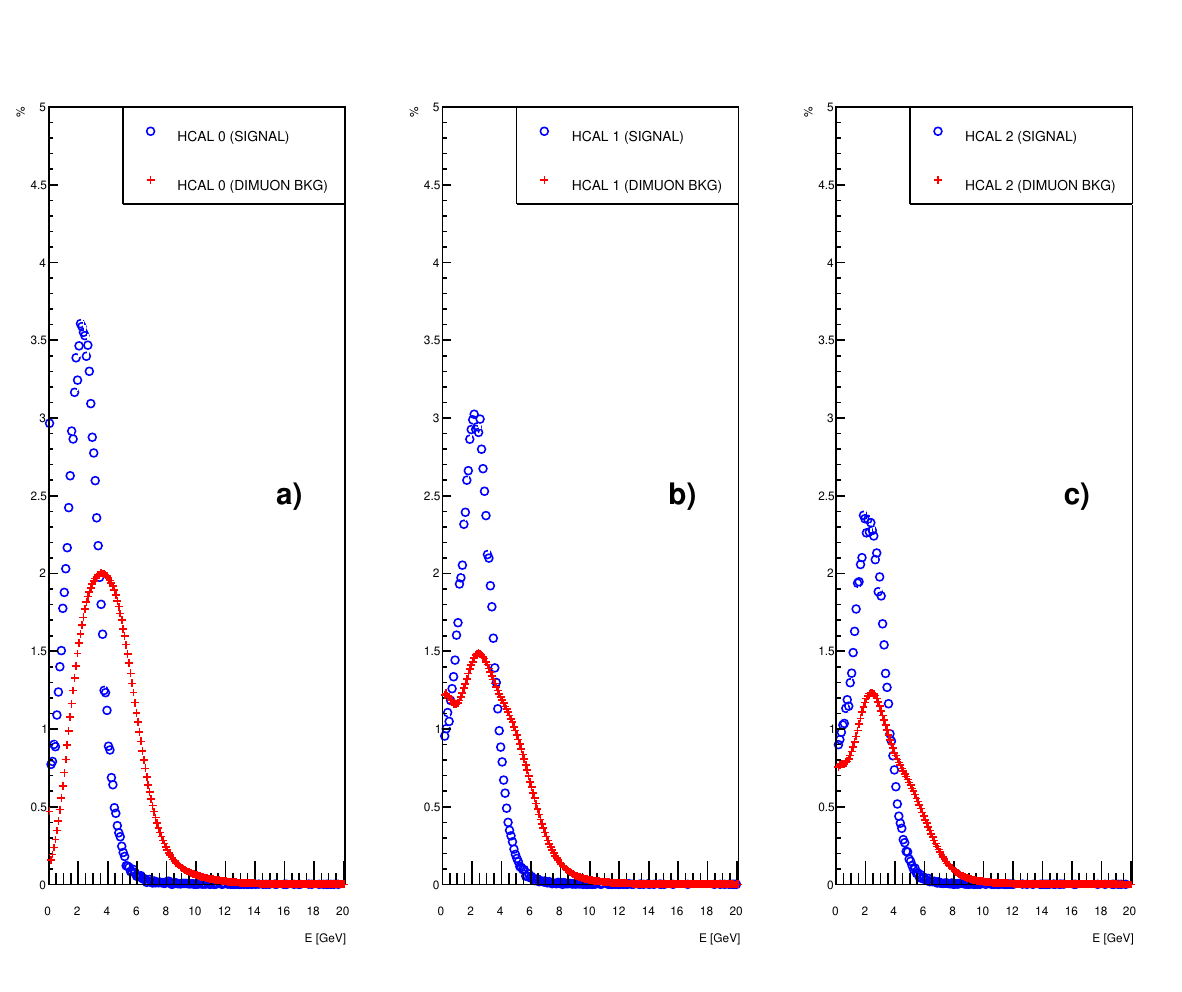}
    \caption{Deposited energy without any selection cuts in each HCAL modules 0-2 $( \mathrm{a})-(\mathrm{c}) $ from simulated signal (blue circle) and dimuon background events (red plus). With increasing HCAL module the signal peak is reduced due to absorption. The background spectrum is likewise reduced and shifted towards a single MIP deposit ($\sim 2.5$ GeV) as one of the dimuons is absorbed.}
    \label{fig:hcaldimuonvssignal}
\end{figure}

\begin{figure}[H]
    \centering
   \includegraphics[width=0.45\textwidth]{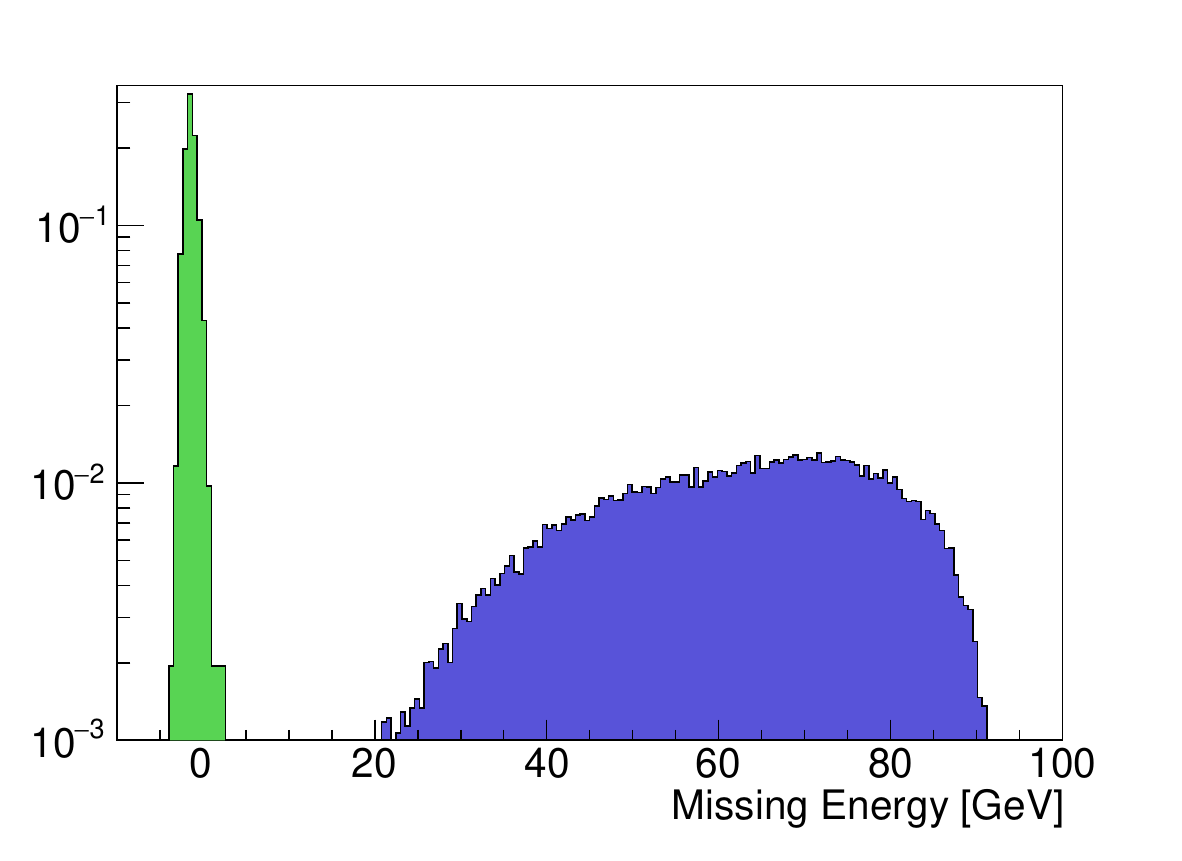}
    \caption{Missing energy $E_{miss} = E_e^i - E_{HCAL} - E_{ECAL} - E_{\mu}^\prime$ for the simulated dimuon background sample (left, green) and signal sample (right, blue) after applying all selection criteria besides missing energy cut. The missing energy for the dimuon background sample is centered around 0, meaning all those events are fully contained in the detector.}
    \label{fig:missing-energy}
\end{figure}

Standard Model events leading to missing energy and a final state muon in the magnet spectrometer can mimic the signal signature. The two dominant background sources in this type of processes are the dimuon production in the target and muonic decays from hadronic beam contamination, $h\rightarrow\mu+X$. Within this work, they have been studied in detail using the MC framework described in the previous section.\\ \indent
Events with dimuon production $\gamma \rightarrow \mu^+ \, \mu^-$ originating from the shower of the primary electron in the ECAL can imitate the signal characteristics if one and only one (anti-)muon of the pair is reconstructed in the magnet spectrometer. The dimuon production at this energy is at the level of $1.05\times10^{-4}$ per electron on target (EOT). The HCAL signature of signal-like events and such background events is shown in Fig. \ref{fig:hcaldimuonvssignal}. The signal spectrum is centered at the single MIP energy deposit of $\sim2.5$ GeV, and is overall reduced as the muons are absorbed deeper into the HCAL setup. The dimuon spectrum displays a larger peak deposit in the first HCAL due to the double MIP deposit, but migrates towards a single MIP spectrum as one of the muons is absorbed.\\ \indent 
Dimuon events where at least one muon is reconstructed in the magnet spectrometer can generally be sorted into three categories: (I) One (anti)muon is absorbed in the ECAL or HCALs while the other is reconstructed in the magnet spectrometer. Such an event is not problematic since the energy is fully contained and thus would not be tagged as a missing energy candidate (II) Both muons pass through the HCALs and are reconstructed in the magnet spectrometer. The requirements (i-iii) on single MIP energy deposit, missing energy, and single track in the magnet remove such events. (III) Both muons penetrate the HCALs but only one is reconstructed in the magnet. These events have missing energy and only one track in the magnet but a double MIP energy deposit in VETO and HCALs, and thus do not pass the selection criteria (ii). \\ \indent
After applying the selection criteria (i-iii), the missing energy variable separates the dimuon background and signal sample as seen in Fig. \ref{fig:missing-energy}. For a simulated background data set equivalent to $1.5 \times 10^{11}$ EOT, the event selection shown in the previous section removes all dimuon events. \\ \indent
The other main background source arises from the muonic in-flight decays of the remaining hadrons in the beam, $h\rightarrow\mu+X$. These events yield a single muon and have also missing energy carried away by the accompanying neutrino. A pion and kaon beam were simulated in the realistic MC simulation framework described in section \ref{sec:search-at-na64} to estimate the background level in these searches considering the hadron admixture quoted in section \ref{sec:search-at-na64} ($\pi/e^-< 10^{-2}$). Muonic in-flight decays occurring before the SRD are suppressed by the NA64 tagging system, so the dangerous events are those occurring between the SRD and the ECAL. In this case, if the muon carries most of the energy, the missing energy cut removes those events. However, events in which the neutrino carries a sizeable missing energy, more than 30 GeV according to simulations, will pass the selection being potential background. After applying the event selection we get a survival rate of $9.6\times 10^{-5}$ for pions and $5.6\times 10^{-5}$ for kaons. The difference in the rates associated with each hadron species is attributed to the underlying decay process. The final states' kinematics from pions and kaons result in different angular acceptance within the magnet spectrometer. From one side, the beam contamination rate of kaons is $3\%$ of that of pions. In addition,  although the probability for kaons to decay between the SRD and ECAL is higher than for pions (0.32\% compared to 0.04\%), the opening angle of the kaon decay products is larger than for pions. For that reason, the momentum reconstruction in the Micromegas located after the bending will further suppress those events. Finally, this result combined with the suppression of the hadron component by the SRD gives a final background event rate of $1.9\times 10^{-11}$ per EOT. 
\section{Sensitivity\label{ref:sensitivity}}
\begin{figure}[H]
    \centering
    \includegraphics[width=0.45\textwidth]{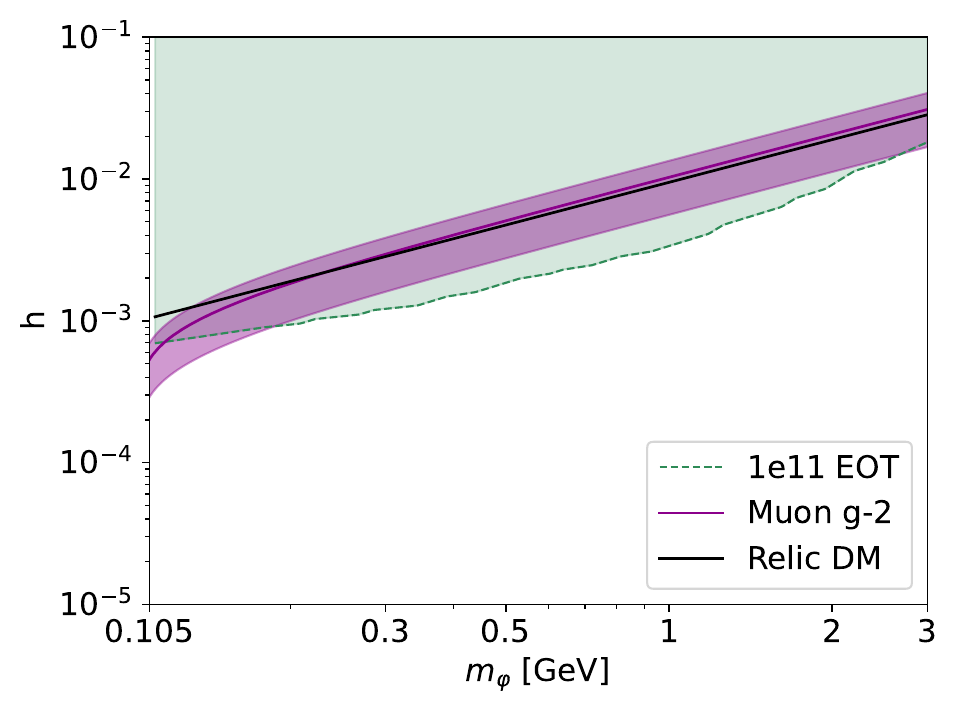}
    \caption{90\% CL projected sensitivity in mass-coupling $(m_\varphi,\ h)$ parameter space of $\varphi$ given the background rate from section \ref{sec:background-sources}. The preferred band to solve the muon $g-2$ anomaly at $3\sigma$ from Eq. \eqref{eq:DLS_g2} is shown in purple. The relic DM annihilation constraint is shown in black and is computed through Eq. \eqref{eq:DM-constraint} with the parameter choices $g_{\varphi \psi_1 \psi_2} = 0.1$ and $m_1 = m_2 = 3m_\varphi$.}
    \label{fig:90cl-gm2}
\end{figure}

The sensitivity of the NA64 experiment to the new leptonic scalar boson $\varphi$ is estimated based on the 90\% confidence level (CL) upper limits that would be obtained under the hypothesis that only SM events are observed. Within this work, it is computed in the modified frequentist approach, using the \texttt{RooFit}/\texttt{RooStats}-based \cite{Verkerke:2003ir,Wolffs:2022fkh,Moneta:2010pm} profile-likelihood ratio statistical test in the asymptotic approximation \cite{Cowan:2010js}. The 90\% CL projected sensitivity in the $(m_\varphi,\ h)$ plane based on the full signal simulation described in section \ref{sec:signal-yield} and considering the background level derived in section \ref{sec:background-sources} is shown in Fig. \ref{fig:90cl-gm2}. With a total statistics of $N_\text{EOT}=10^{11}$, NA64 has the reach to probe a broad region of the parameter space compatible with a fermionic DM relic abundance (see Eq. \ref{eq:DM-constraint}) for masses above the muon mass up to few GeV. Additionally, it can also provide stringent constraints on the allowed $(m_\varphi,\ h)$ values in the explanation of the muon $g-2$ up to masses $m_\varphi\lesssim2m_\mu\sim\mathcal{O}(200\ \text{MeV})$ and coupling $h\lesssim7\times10^{-4}$.
\section{Conclusion\label{sec:conclusion}}
In this paper, we presented the capabilities of fixed target experiments such as NA64, to search for processes involving electron-to-muon transitions. In particular, in this study, we have focused on using the 100 GeV electron beam at H4 to search for a new lepton-flavor-changing scalar boson in the Bremsstrahlung-like reaction $e N\rightarrow \mu N \varphi$. We have calculated the exact-tree level differential and total cross-sections involved in such a process and its WW phase-space approximation. The corresponding formulas have been implemented in a realistic Monte Carlo framework using the \texttt{GEANT4}-based \texttt{DMG4} package as well as the NA64 experimental setup. The two main potential background sources, dimuons production in the ECAL and the remaining hadron decay in-flight have also been identified and studied. We have shown that the most dangerous background arises from the remaining muons from the beam hadrons decay-in-flight. This could be reduced further by improvements of the SRD which are currently under study. In this paper, we have demonstrated that adding a spectrometer after the last HCAL module to reconstruct the final state muon momentum will allow to suppress the background at the required level. The experiment projected sensitivity based on simulations indicates that with $10^{11}$ EOT NA64 can probe the parameter space motivated by a simultaneous solution of the muon g-2 anomaly and the DM problem for $0.2 \leq m_\varphi \leq 1$ GeV. The same process can be also probed in the NA64 muon configuration searching for $\mu\rightarrow e$ and will be the subject of future works. In addition, the positrons resulting from the electromagnetic shower initiated by the incoming electron can also induce the reaction $e^+ N\rightarrow \mu^+ N\varphi$. In this case, a signal-like event will be identified as detecting an incoming $e^-$ and an outgoing $\mu^+$. This study will also be covered in future works. Finally, this work opens the possibility to further exploit other processes involving electron-to-muon conversion using the NA64 missing energy and momentum techniques as ALP \cite{Jho:2022snj}, scalar\cite{Araki:2021vhy} or dark vector \cite{Zhevlakov:2023jzt, Heeck:2016xkh}.

\section{Acknowledgments}
We acknowledge the members of the NA64 collaboration for fruitful discussions. In particular, we thank N. V. Krasnikov for his helpful suggestions and correspondence. The work of P. Crivelli, B. Banto Oberhauser, M. Mongillo, and H. Sieber is supported by the SNSF Grants  No. 186158, and No. 197346 No. 216602, No.  219485(Switzerland) and by ETH Grant No. 22-2 ETH-031. The work of L. Molina-Bueno is supported by SNSF Grant No. 186158 (Switzerland), RyC-030551-I, PID2021-123955NA-100 and CNS2022-135850 funded by MCIN/AEI/ 10.13039/501100011033/FEDER, UE (Spain).
\appendix
\section{Improved Weisz\"{a}cker-Williams approach\label{app:iww-approach}}
The computations performed in section \ref{sec:ww-approximation} can be further simplified under the assumption that the lower bound of the effective photon flux is independent of $x$ and $\theta_k$. As such we can pull $\chi^\text{WW} \rightarrow \chi^\text{IWW}$ outside of the integral over $\cos{\theta_{k}}$ in Eq. \eqref{eq:dsdx_WW}, permitting us to derive a fully analytical differential cross section. The new bounds of $\chi^\text{IWW}$ are $t_\text{min} = m_{\varphi}^4/(4 E_{i}^2)$ and $t_\text{max} = m_{\varphi}^2 + m_{i}^2$. The integral over $\theta_k$ in Eq. \eqref{eq:dsdx_WW} is analytically calculated by changing variables from $\cos{\theta_{k}}$ to $\tilde{u}$, where we use $\tilde{u}$ evaluated at $t_\text{min}$ just like in the WW approximation. This allows us to expand $ \sin{\theta_{k}} = \theta_{k} + \mathcal{O}(\theta_{k}^3)$ leading to $d \cos{\theta_{k}} \simeq - \theta_{k} \cdot d\theta_{k}$ and thus $d\tilde{u}  = 2x E_{i}^2 d \cos{\theta_{k}}$. After the substitution, the single-differential cross-section takes the following form
\begin{equation}
    \label{eq:dsdx_IWW}
    \left( \frac{d \sigma}{dx}\right)_\text{IWW}  = \frac{\alpha^{2}h^2 }{4 \pi} \beta_{k} (1-x) \chi^\text{IWW} \int_{\tilde{u}_\text{min}}^{\tilde{u}_\text{max}} \,d \tilde{u}\ \frac{|\mathcal{A}_\varphi^{2 \rightarrow 2}|^{2}_{t_\text{min}}}{\tilde{u}^2} ,
\end{equation}
where $\tilde{u}_\text{max} = \tilde{u}(0)$ and $\tilde{u}_\text{min} = \tilde{u}(\theta_\text{max})$. Using the expression for $|\mathcal{A}_\varphi^{2 \rightarrow 2}|^2$ in Eq. \eqref{eq:a2tmin} the integral in Eq. \eqref{eq:dsdx_IWW} can finally be solved analytically as
\begin{equation}
    \begin{split}
    \int_{\tilde{u}_\text{min}}^{\tilde{u}_\text{max}} &\,d \tilde{u}\ \frac{|\mathcal{A}_\varphi^{2 \rightarrow 2}|^{2}_{t_\text{min}}}{\tilde{u}^2} = \Bigg \{ \frac{x^2}{-2(1-x)\tilde{u}} + \frac{\Delta m^2x}{-2\tilde{u}^2} \\ 
    &+\Delta m^2 \frac{ m_{\varphi}^2 (1-x) + x [ m_{f}^2+m_{i}^2(x-1) ] }{-3\tilde{u}^3} \Bigg\}\biggr\rvert_{\tilde{u}_\text{min}}^{\tilde{u}_\text{max}}. 
    \end{split}
\end{equation}
Similarly, the IWW approximation can be applied to the single-differential cross-section with respect to the fractional lepton energy, $y$. Integrating Eq. \eqref{eq:dsdydpsi_WW} over $\cos{\psi}$ gives the differential cross-section\\
\begin{equation}
    \begin{split}
    \left( \frac{d \sigma}{d y} \right)_\text{IWW} &= \frac{\alpha^2 h^2}{2\pi} E_{i}^2 \beta_{f} \chi^\text{IWW} (1-y)^3 \int_{0}^{\psi_\text{max}} \, d \cos{\psi} \ \\ \times  \bigg\{ \frac{1}{2 \tilde{t}^2} &- \frac{\Delta m^2}{\tilde{t}^3} + \frac{\Delta m^2 [ m_{f}^2 + y ( m_{i}^2 y + \Delta m^2) ] }{y \tilde{t}^4}  \bigg\}.
    \end{split}
    \label{eq:dsdy_IWW_with_integral}
\end{equation}
As in section \ref{sec:ww-approximation}, the maximal $\psi$ angle is estimated through numerical integration and bound dependence on the contribution to the integral. It is found $\psi_{max} = 0.3$. We thus perform the integral in Eq. \ref{eq:dsdy_IWW_with_integral} analytically through the change of variables from $ \cos{\psi} \rightarrow \tilde{t}$. Since we use the IWW approximation we can assume $d \cos{\psi} \simeq - \psi d \psi$. The Jacobian of the transformation is $d\tilde{t} \simeq 2y E_{i}^2 \psi d \psi-2y E_{i}^2 d \cos{\psi}$. Eq. \eqref{eq:dsdy_IWW_with_integral} becomes
\begin{equation}
    \begin{split}
    \left( \frac{d \sigma}{d y} \right)_\text{IWW} &= \frac{\alpha^2 h^2}{4\pi} \beta_{f} \chi^\text{IWW} \frac{(1-y)^3}{y} \\ 
    \times \Bigg\{ \frac{-1}{2 \tilde{t}} &+ \frac{\Delta m^2}{2 \tilde{t}^2} - \frac{\Delta m^2 [ m_{f}^2 + y ( m_{i}^2 y + \Delta m^2) ] }{3 y\tilde{t}^3}  \Bigg\}\biggr\rvert_{\tilde{t}_\text{min}}^{\tilde{t}_\text{max}},
    \end{split}
\end{equation}
with $t_\text{min}$ and $t_\text{max}$ defined previously. For completeness, it is worth noting that the IWW single-differential cross-section for the outgoing lepton scattered angle is given by integrating Eq. \eqref{eq:dsdydpsi_WW} over \textit{y}.
\bibliography{bibl}	
\end{document}